\documentclass[review,12pt,authoryear]{elsarticle}
\usepackage{amssymb}
\usepackage{amsthm}
\usepackage{amsmath}
\usepackage{mathrsfs}
\usepackage{graphicx}
\usepackage{epstopdf}
\usepackage{float}
\usepackage{caption}
\usepackage{soul}
\usepackage{color,soul} %to highlight text
\usepackage{color} %To highlight references in the text
\usepackage{subcaption}
\usepackage{bm}
\usepackage{bbm}
\usepackage{mathrsfs}
\usepackage{cleveref}
\usepackage{soul}
\usepackage[margin=3cm]{geometry}% to reduce margins
\usepackage{booktabs}
\usepackage{chngpage}
\journal{Journal of Natural Gas Science and Engineering}

\makeatletter
\def\@author#1{\g@addto@macro\elsauthors{\normalsize%
    \def\baselinestretch{1}%
    \upshape\authorsep#1\unskip\textsuperscript{%
      \ifx\@fnmark\@empty\else\unskip\sep\@fnmark\let\sep=,\fi
      \ifx\@corref\@empty\else\unskip\sep\@corref\let\sep=,\fi
      }%
    \def\authorsep{\unskip,\space}%
    \global\let\@fnmark\@empty
    \global\let\@corref\@empty  %% Added
    \global\let\sep\@empty}%
    \@eadauthor={#1}
}
\makeatother

\makeatletter
\def\thickhline{%
  \noalign{\ifnum0=`}\fi\hrule \@height \thickarrayrulewidth \futurelet
   \reserved@a\@xthickhline}
\def\@xthickhline{\ifx\reserved@a\thickhline
               \vskip\doublerulesep
               \vskip-\thickarrayrulewidth
             \fi
      \ifnum0=`{\fi}}
\makeatother

\newlength{\thickarrayrulewidth}
\begin{document}

\begin{frontmatter}

%% Title, authors and addresses

%% use the tnoteref command within \title for footnotes;
%% use the tnotetext command for theassociated footnote;
%% use the fnref command within \author or \address for footnotes;
%% use the fntext command for theassociated footnote;
%% use the corref command within \author for corresponding author footnotes;
%% use the cortext command for theassociated footnote;
%% use the ead command for the email address,
%% and the form \ead[url] for the home page:
%% \title{Title\tnoteref{label1}}
%% \tnotetext[label1]{}
%% \author{Name\corref{cor1}\fnref{label2}}
%% \ead{email address}
%% \ead[url]{home page}
%% \fntext[label2]{}
%% \cortext[cor1]{}
%% \address{Address\fnref{label3}}
%% \fntext[label3]{}

\title{Comparison of hydrogen diffusivities measured by electrochemical permeation and temperature-programmed desorption in cold-rolled pure iron}

%% use optional labels to link authors explicitly to addresses:
%% \author[label1,label2]{}
%% \address[label1]{}
%% \address[label2]{}

\author{Alfredo Zafra\fnref{IC}}

\author{Zachary Harris\fnref{UVa}}

\author{Chao Sun\fnref{IC}}

\author{Emilio Mart\'{\i}nez-Pa\~neda\corref{cor1}\fnref{IC}}
\ead{e.martinez-paneda@imperial.ac.uk}

\address[IC]{Department of Civil and Environmental Engineering, Imperial College London, London SW7 2AZ, UK}

\address[UVa]{Department of Materials Science and Engineering, University of Virginia, Charlottesville, VA 22904, USA}

\cortext[cor1]{Corresponding author.}

\begin{abstract}
The diffusivity of hydrogen in cold-rolled pure iron is investigated using permeation and desorption methods. Electrochemical charging, electro-permeation and thermal desorption spectroscopy (TDS) experiments are conducted. Firstly, the relation between the charging current and the hydrogen concentration is established. Secondly, permeation experiments are conducted at 22, 40 and 67$^\circ$C to quantify the diffusivity dependence on temperature. Finally, the diffusivity is estimated by using two types of desorption experiments and Fick's law: (i) a `rest time' method, by which we measure the hydrogen content of samples held at room temperature for different times, and (ii) isothermal desorption experiments at temperatures ranging from 22 to 80$^\circ$C, fitting the resulting desorption rate versus time curves. Good agreement is obtained between the isothermal desorption and permeation approaches, with observed differences discussed and rationalised. Moreover, measured diffusivity values for cold-rolled pure iron are also found to be comparable to those reported in the literature. This work demonstrates that isothermal desorption experiments are a convenient approach to determine hydrogen diffusivity over a wide range of temperatures, as facilitated by new TDS systems with fast heating rates. 
\end{abstract}

\begin{keyword}

Hydrogen \sep Electro-permeation \sep Diffusion \sep Thermal Desorption Spectroscopy  \sep Isothermal desorption
%% keywords here, in the form: keyword \sep keyword

%% PACS codes here, in the form: \PACS code \sep code

%% MSC codes here, in the form: \MSC code \sep code
%% or \MSC[2008] code \sep code (2000 is the default)

\end{keyword}

\end{frontmatter}

%% \linenumbers

%% main text
\section{Introduction}
\label{Introduction}

Hydrogen-induced degradation is a pertinent life-limiting damage mode for many metallic structural components across the aerospace, marine, energy, transportation, and infrastructure sectors \citep{Gangloff2003}. Critically, the presence of hydrogen-assisted sub-critical cracking compromises structural integrity management approaches, thereby complicating life prediction and fracture control efforts \citep{Gangloff2003,Gangloff2016}. Such effects become increasingly important as novel approaches to increasing the viability of hydrogen energy technologies are considered \citep{Gangloff2012}, which may lead to existing infrastructure being subjected to unexpected operating conditions. For example, feasibility studies have suggested that blending hydrogen into natural gas pipeline networks is one pathway by which gaseous hydrogen may be economically stored and transported \citep{Melaina2013,Hafsi2019,Ishaq2020}. However, recent experimental evaluations have also demonstrated that the exposure of pipeline steels to natural gas/hydrogen mixtures can result in accelerated fatigue crack growth rates \citep{Meng2017,Shang2020}, degraded fracture resistance \citep{Nguyen2020}, and reduced tensile properties (\textit{e.g.}, breaking stress, notch tensile strength, ductility, etc.) \citep{Meng2017}. Critically, the extent of the degradation in mechanical properties is sensitive to the hydrogen volume fraction in the natural gas/hydrogen mixture, suggesting a fundamental dependence of the degradation mechanism on the available hydrogen content \citep{Meng2017}.

Driven by this deleterious impact on performance, over a century of scientific study \citep{Johnson1875} has sought to mechanistically understand the microscale processes by which hydrogen degrades the mechanical properties of structural metals. These efforts have led to the development of numerous theories \citep{Robertson2015,Gerberich2012,Shishvan2020}, but a growing literature database strongly suggests that hydrogen-induced degradation proceeds \textit{via} the synergistic interaction of several different hydrogen-modified processes \citep{Gangloff2017,Djukic2019,Robertson2015}; \textit{i.e.}, hydrogen-induced reductions in grain boundary cohesive strength \citep{Harris2018,JMPS2020}, hydrogen-modified plasticity behavior \citep{Barnoush2010a,Nagao2018,Wang2014,Harris2020c}, hydrogen-induced stabilization of vacancies \citep{Lawrence2017,Nagumo2019}, \textit{etc.}. Critically, the relative contributions of these various hydrogen-modified processes to material degradation are likely to evolve with changes in hydrogen content \citep{Harris2018}. As such, in addition to evaluating the operative hydrogen degradation mechanisms, it is equally important to develop a clear understanding of fundamental hydrogen-material interactions (\textit{e.g.}, uptake, trapping, and diffusion) \citep{Depover2021,Gangloff2017,CS2020b}. It is well-established that differences in hydrogen-material interactions alone can modify susceptibility to hydrogen-induced degradation \citep{Nelson1973}. For example, the magnitude of the Stage II crack growth rate commonly observed during hydrogen environment-assisted cracking in severe environments has been shown to linearly scale with differences in hydrogen diffusivity \citep{Gangloff2008,AM2016,Harris2021}. Similarly, the threshold stress intensity for hydrogen-assisted cracking generally correlates well with the diffusible hydrogen concentration \citep{Akhurst1981,Gangloff2017,CMAME2018}. Lastly, for closed systems containing a finite hydrogen concentration, the introduction of a uniform distribution of strong hydrogen trap sites (such as carbides) has been demonstrated to reduce susceptibility \textit{via} the sequestering of hydrogen into these benign locations \citep{Bhadeshia2016,AM2020}.

Several different experimental approaches exist for quantifying hydrogen-metal interactions. For example, inert gas fusion methods are commonly employed to determine the total hydrogen content \citep{Lawrenz2006}, while the barnacle cell electrode method can be used to quantify the diffusible hydrogen content under electrochemical charging conditions \citep{DeLuccia1981}. However, the two most common methods for assessing hydrogen-metal interactions are permeation and thermal desorption \citep{Depover2021}. Briefly, permeation experiments are conducted by separating two independent environments by means of a thin membrane made of the material of interest. Atomic hydrogen is generated and absorbed on one side of the membrane, it then diffuses through the membrane, and subsequently effuses out the other side. Permeation can be conducted using either gaseous or electrochemically generated hydrogen, with the former being common for measuring permeation at elevated temperatures \citep{Johnson1988}. For gaseous permeation, it is typical for the egress side of the membrane to be held under vacuum, so that the permeation rate of hydrogen can be monitored via mass spectrometer \citep{Choi1970} or the change in vacuum pressure \citep{Stross1956}. For electrochemical permeation, the permeation rate is measured via the current induced by the oxidation of hydrogen as it effuses from the membrane \citep{Subramanyan1981,Boes1976}. Conversely, thermal desorption-based methods typically use specimens that have been precharged with hydrogen, which are placed into a vacuum chamber and heated according to a pre-programmed temperature versus time profile \citep{Verbeken2012}. The hydrogen desorbing from the specimen is then measured by a mass spectrometer. Post-experiment analysis of the obtained permeation rate or desorbed hydrogen content versus time (or temperature) data is then conducted to determine information on hydrogen-metal interactions, such as hydrogen solubility, trapping, and diffusivity behavior \citep{Galindo-Nava2017a,IJF2020,IJHE2020,TAFM2020b}.

Given that desorption-based methods are predominantly performed on precharged specimens, the necessity of conducting the experiment under high vacuum conditions limits the utility of this approach for materials with a `fast' hydrogen diffusivity (such as pure Fe). Specifically, while it is certainly possible to obtain information on the trapping behavior of such materials with desorption approaches \citep{Choo1982,Choo1983,Lee1987}, the rapid egress of the lattice hydrogen can obfuscate assessments of hydrogen solubility and diffusivity. As such, it is common for `fast' diffusing materials to be characterized using permeation methods. However, permeation experiments can exhibit substantial scatter. For example, hydrogen diffusivities determined from permeation experiments for nominally pure Fe at ambient temperature (25$^\circ$C) can differ by multiple orders of magnitude across studies \citep{Kumnick1974}. This significant variability in permeation-measured data has been attributed to a wide range of potential influences, including: testing variable sensitivities \citep{Gonzalez1969,Turnbull1995}, surface effects \citep{Kiuchi1983,Addach2009}, nonsteady-state conditions \citep{Nelson1973}, concentration-dependent diffusion \citep{Ono1968,Zafra2020}, hydrogen trapping \citep{Kumnick1974,Oriani1970}, and analysis method \citep{Boes1976,Carvalho2017}. 

These aforementioned challenges associated with assessing the diffusivity of hydrogen in pure Fe using permeation experiments strongly motivate exploring the use of desorption-based approaches for `fast' diffusing alloys. For example, isothermal desorption experiments have been used to generate hydrogen diffusivity versus temperature relationships in stainless steels \citep{Mine2009}, Ni-based superalloys \citep{Ai2013}, Cu-Ni and Ni-Cr \citep{Matsuo2014}, a number of low-alloy steels \citep{Yamabe2015}, and precipitation-hardened steels \citep{Yamabe2021}. Critically, in order to use this isothermal method for low-alloy steels, large specimens up to 19 mm in diameter were required \citep{Yamabe2015}. Recent advances in thermal desorption equipment, including improved conduction-based heating capabilities that can induce controlled heating rates of up to 60$^\circ$C/min, vacuum systems with dedicated loading chambers for fast evacuation to reduce sample `rest time', and improved resolution mass spectrometers, suggest that this desorption-based approach could be used to obtain near-ambient ($<100^\circ$C) hydrogen diffusivity data in `fast' diffusing materials, such as pure iron, without the need for large specimens. However, such an evaluation and the follow-on comparison between isothermal desorption and permeation-determined diffusivities have yet to be performed. 

The objective of this study is to compare permeation and thermal desorption-based assessments of hydrogen diffusivity using 1-mm thick cold-rolled pure iron to establish the efficacy of desorption-based approaches for measuring hydrogen diffusivities in a representative `fast' diffusing material. The total hydrogen concentration versus applied current density relationship is first established for cold-rolled pure iron to inform the current density to be employed in subsequent experiments. Permeation measurements are performed at temperatures ranging from 22 to 67$^\circ$C and the hydrogen diffusivity is then estimated using three different approaches: breakthrough, lag time, and by fitting the permeation transient to an approximate solution of Fick's law. These data are then compared to hydrogen diffusivity values determined from both variable rest time and isothermal (ranging from 22 to 80$^\circ$C) desorption experiments. We extensively discuss differences and similarities, as well as the implications for diffusion data extraction. 

\section{Experimental Methods}
\label{Sec:Expt}

\subsection{Material}
This study was conducted using cold-rolled pure iron (supplier-reported purity of $>$99.5 wt. \%Fe) procured in the as-rolled condition from Goodfellow Ltd. as a 1-mm thick sheet. The supplier-reported average degree of cold work was 50\%. All experiments were performed on thin plate specimens with nominal dimensions of 250 mm x 250 mm x 1 mm (electropermeation) and 10 mm x 10 mm x 1 mm (desorption), which were excised from the sheet using an abrasive saw. Each face of the sample was iteratively ground flat using SiC papers, finishing at 1200 grit. 

\subsection{Electropermeation}
\label{Subsec:Electropermeation}
\begin{figure}[H]
     \centering
         \centering
         \includegraphics[width=1\textwidth]{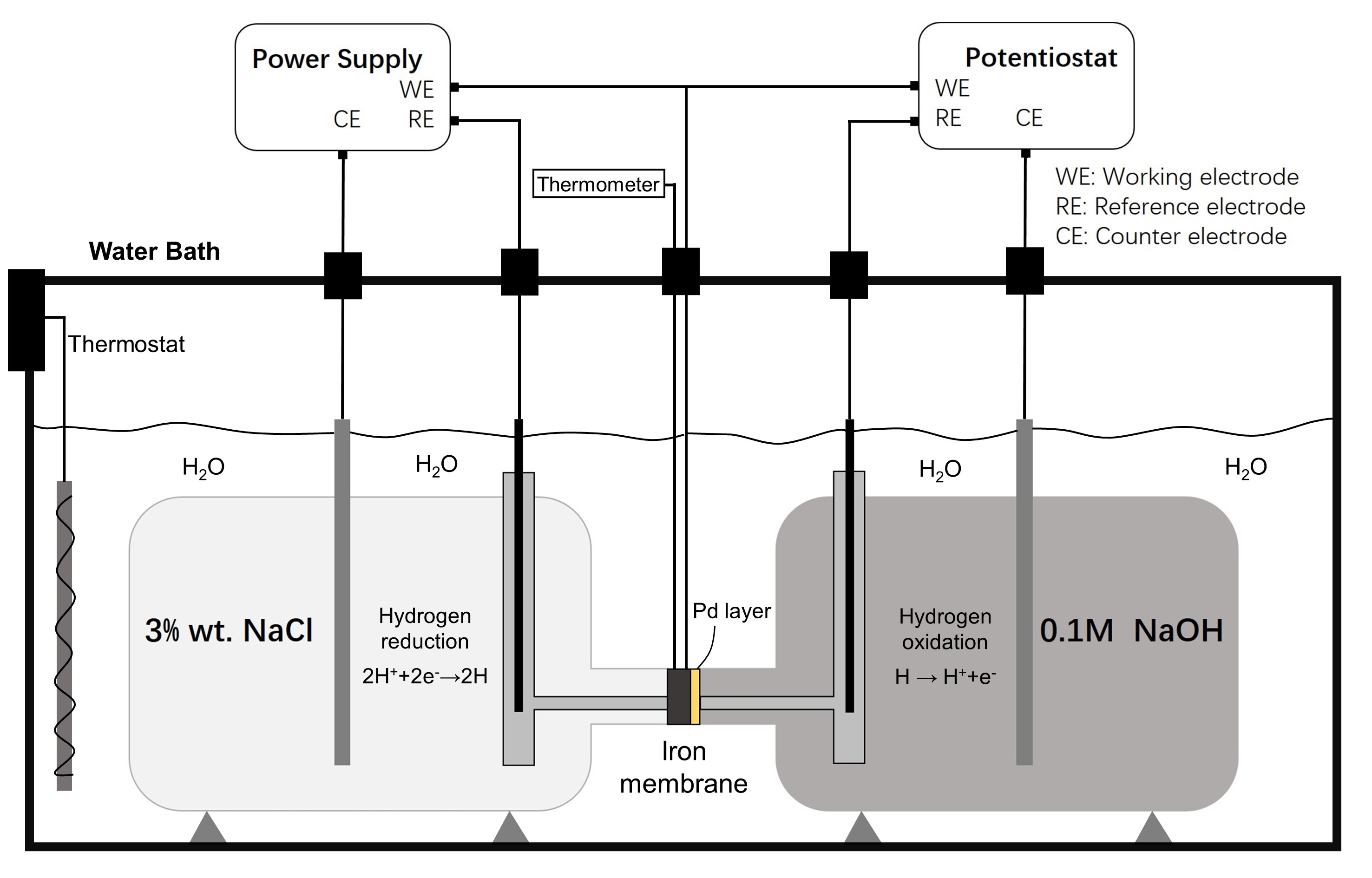}
        \caption{Schematic of the electrochemical permeation testing configuration used in this work. The permeation setup is placed into a water bath to conduct high-temperature experiments.}
        \label{fig:PermeationCell}
\end{figure}

Electrochemical permeation experiments were performed at temperatures of 22, 40, and 67$^\circ$C using a modified Devanathan-Stachurski double-cell that was capable of being submerged in a hot bath; a schematic of this system is shown in Fig. \ref{fig:PermeationCell}. Two tests were conducted for each temperature. A masking plate was used to consistently expose a circular area of 2 cm$^2$ (16-mm diameter) on the sample membrane to both sides of the double-cell. For the experiments conducted at 40 and 67$^\circ$C, the hot bath temperature was actively regulated by a thermostat and continuously monitored with a thermometer, which was in contact with the specimen. This setup resulted in a variation of $\pm1^\circ$C in temperature over the course of a given permeation experiment. Note that the temperature of the solution at each side of the double-cell was allowed to stabilize prior to the start of each elevated temperature permeation experiment. The three-electrode hydrogen reduction cell was filled with 3 wt. \% NaCl solution and contained a Pt counter electrode and a saturated calomel reference electrode. Hydrogen production was achieved by applying a current density of 5 mA/cm$^2$ to the cold-rolled Fe membrane using a Gamry 1010B potentiostat operated in galvanostatic mode. The hydrogen oxidation cell was filled with 0.1 M NaOH solution and also contained a Pt counter electrode and a saturated calomel reference electrode, with a second Gamry 1010B potentiostat operated in chronoamperometry mode to record the hydrogen permeation current density, $J_p$, as a function of time. The hydrogen reduction and oxidation reactions taking place at the entry and exit surfaces of the specimen, respectively, are also included in Fig. 1.

Prior to conducting each permeation experiment, a thin layer of Pd (approximately 50-nm thick) was electroplated onto the specimen surface facing the oxidation side of the double-cell. Electroplating was completed by applying a current density of 1 mA/cm$^2$ to the specimen for approximately 5 minutes while it was immersed in a commercial solution that contained 2 g/L of Pd. The Pd layer is employed for two reasons \citep{Manolatos1995}: (1) to enhance the hydrogen oxidation reaction kinetics so as to minimize the hydrogen concentration at the sample surface in the oxidation cell, and (2) to avoid the oxidation of iron, which would obfuscate the true hydrogen permeation current. Once plated, the specimen was inserted into the Devanathan-Stachurski double-cell and the open-circuit potential (OCP) of the sample membrane was monitored on the oxidation side of the cell for 1 hour. Upon completing this 1 hour hold, the oxidation side of the membrane was then polarized to the final potential recorded during the OCP measurement (typically between -40 and -60 mV$_{SCE}$). The permeation current density was then allowed to stabilize to a value less than 0.1 to 0.2 $\mu$A/cm$^2$ before the galvanostatic cathodic charging was started on the reduction side of the double-cell.

\subsubsection{Determination of the diffusion coefficient}
\begin{figure}[H]
     \centering
         \centering
         \includegraphics[width=0.9\textwidth]{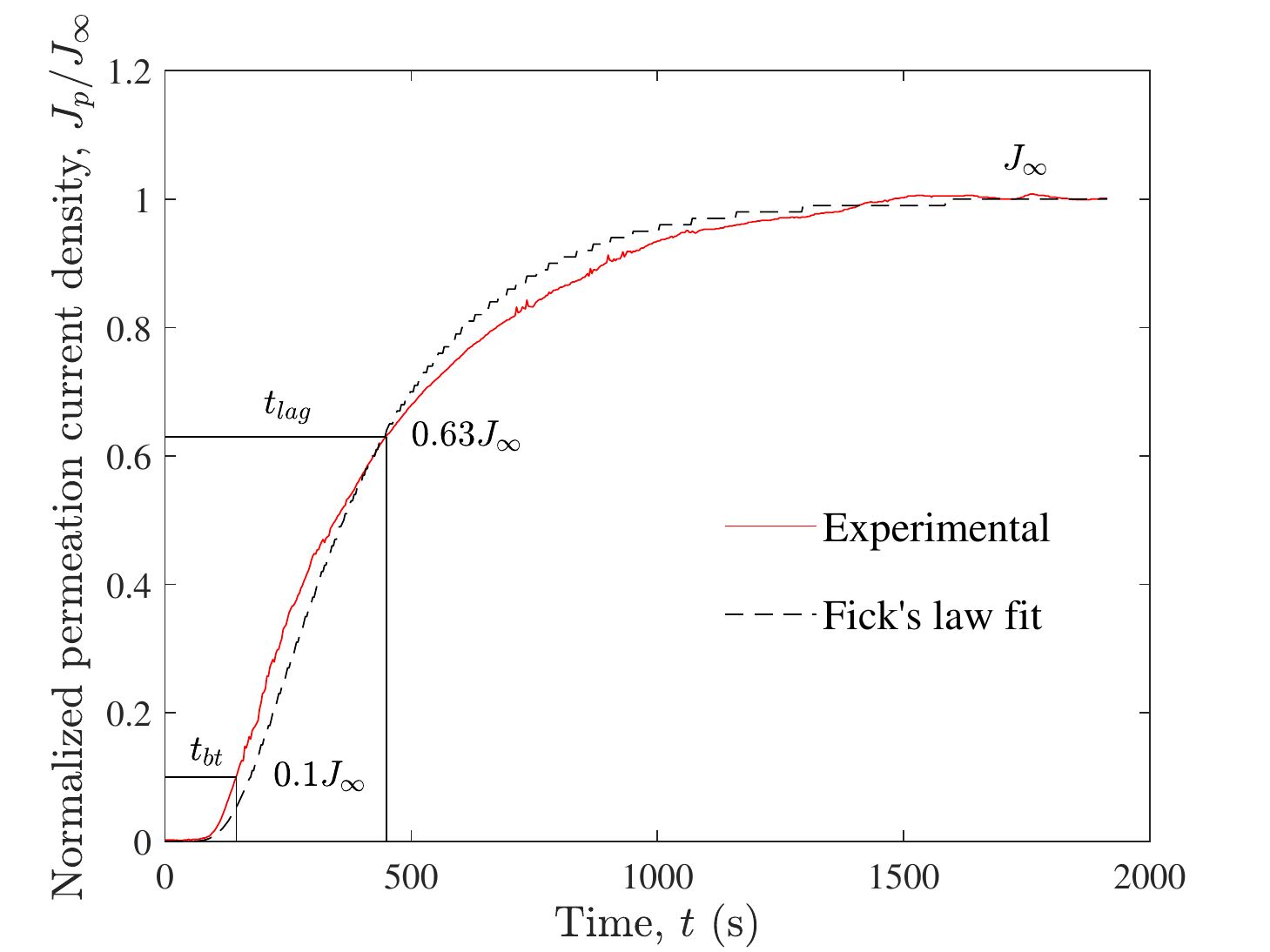}
        \caption{Typical rise permeation transient, with the three main methods used to determine the diffusivity superimposed on the experimental results.}
        \label{fig:PermeCurvesExperimentalProc}
\end{figure}

A representative hydrogen permeation transient versus time relationship that would be generated during an electrochemical permeation experiment is provided in Fig. \ref{fig:PermeCurvesExperimentalProc}. From these data, the hydrogen diffusivity can be calculated using three different methods \citep{Turnbull1989}: (i) the breakthrough time method, (ii) the lag time method, and (iii) by fitting the permeation transient to an approximate solution of Fick's law \citep{Crank1979}. The last case involves fitting the entire permeation transient, while the first two approaches use closed-form solutions to relate the time required to reach specific fractions of the steady state permeation current density, $J_\infty$, to the hydrogen diffusion coefficient $D$. In each case, it is assumed that the hydrogen subsurface concentration is a constant finite value at the entry side and zero at the exit side of the membrane.

Regarding the first method, the breakthrough time, $t_{bt}$, is identified as the time required for  $J_p$ to reach 10\% of $J_\infty$ (\emph{i.e.}, $J_p$/$J_\infty$=0.1) and is nominally considered to be the time required for the first hydrogen atoms to permeate completely through the membrane. Assuming a membrane geometry of thickness $L$, the hydrogen diffusivity, $D_{bt}$, can then be analytically determined for a given $t_{bt}$ using:
\begin{equation}\label{eq:Dbt}
    D_{bt} = \frac{L^2}{15.3t_{bt}} 
\end{equation}

For the lag time method, the hydrogen diffusivity, $D_{lag}$, is obtained in a similar manner as the breakthrough approach. Specifically, the lag time, $t_{lag}$, is identified as the time required to achieve a $J_p$ that is 63\% of $J_\infty$, which is then related to $D_{lag}$ using:
\begin{equation}\label{eq:Dlag}
    D_{lag} = \frac{L^2}{6t_{lag}}
\end{equation}
Lastly, the entire permeation transient as a function of time can be fit \emph{via} least-squares regression to an approximate solution to Fick's second law obtained from either a Laplace or Fourier transform \citep{Crank1979}, assuming 1-D diffusion. For this approach, the hydrogen diffusivity, $D_{lpc}$, is used as the sole fitting parameter to achieve the best fit between the experiment and predicted $J_p$/$J_\infty$ versus time relationship, with the predicted relationship determined from:
\begin{equation}\label{eq:Dlpc}
    \frac{J_p}{J_\infty}= \frac{2}{\pi^{1/2}}\frac{L}{(D_{lpc} t)^{1/2}}\exp\left(-\frac{L^2}{4D_{lpc}t}\right)
\end{equation}

The hydrogen diffusion coefficients measured with these three techniques were compared and differences discussed. 

\subsection{Hydrogen charging}
\label{Sec:Hcharging}

\begin{figure}[H]
     \centering
         \centering
         \includegraphics[width=0.85\textwidth]{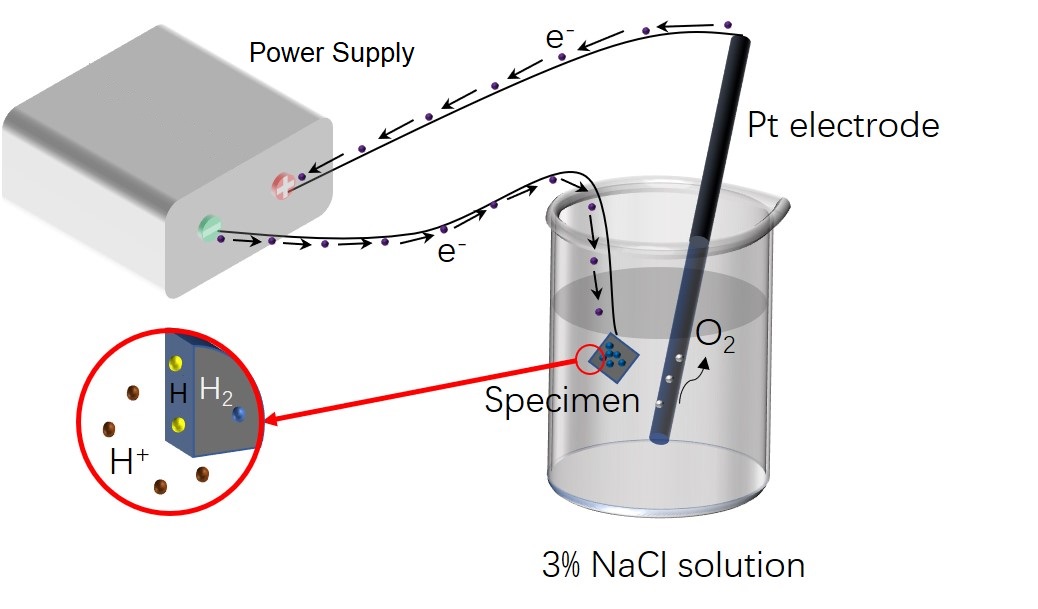}
        \caption{Schematic of the electrochemical hydrogen charging setup.}
        \label{fig:CellCathodicCharge}
\end{figure}

Specimens were electrochemically precharged with hydrogen using a Gamry 1010B potentiostat operated in galvanostatic mode to maintain a constant cathodic current density, $J_c$. All charging experiments were performed with the specimen fully immersed in 3 wt. \% NaCl solution and referenced to a Pt electrode; a schematic of the hydrogen charging setup is shown in Fig. \ref{fig:CellCathodicCharge}. Based on previously reported diffusivities for cold-rolled Fe ($D \approx 8\times10^{-11}$ m$^2$/s; \citealp{VandenEeckhout2017}), all specimens were precharged for 3 hours to obtain a nominally uniform hydrogen concentration across the plate thickness. 

\subsection{TDS measurements}

\begin{figure}[H]
     \centering
         \centering
         \includegraphics[width=0.8\textwidth]{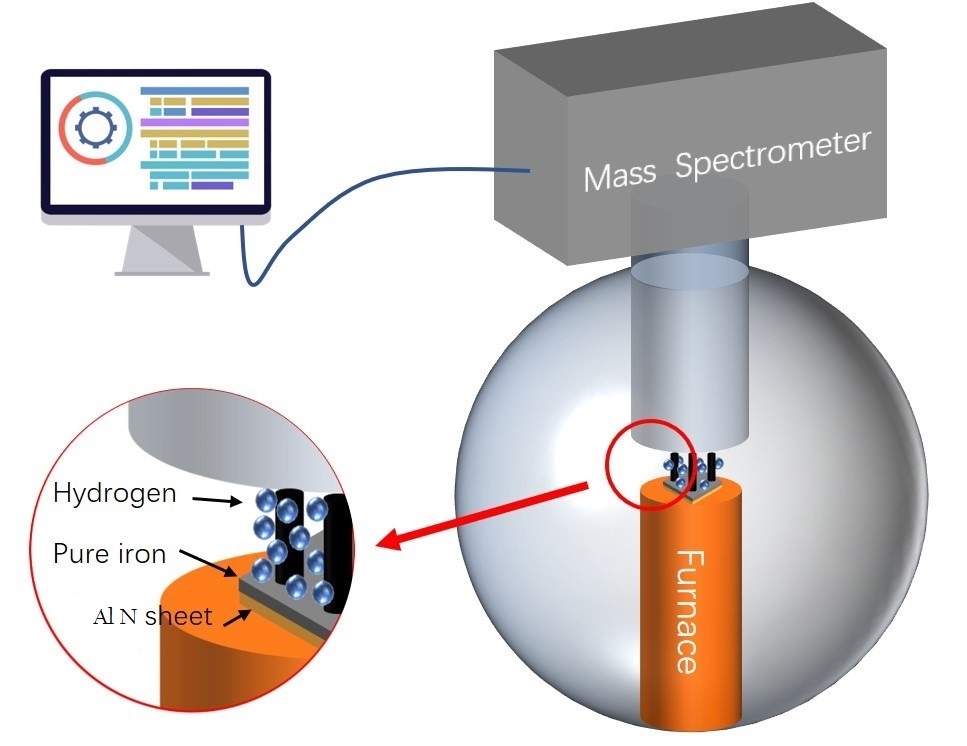}
        \caption{Schematic representation of the thermal desorption spectrometer during a measurement, including a detailed sketch of sample fixation system used to maximize the heating response.}
        \label{fig:TDSSetup}
\end{figure}

All desorption-based experiments were completed using a dedicated thermal desorption spectroscopy (TDS) system capable of achieving heat rates as fast as 60$^\circ$C/min (3600 K/h). The system consisted of a dedicated analysis chamber, which was maintained at a vacuum pressure of 10$^{-9}$ mbar using a turbomolecular pump, and a small sample loading chamber designed to minimize the time required to reach vacuum levels comparable to the analysis chamber. Hydrogen content measurements were made using a regularly calibrated Hiden Analytical RC PIC quadrupole mass spectrometer, which had a detection resolution of $4.4\times10^{-6}$ wppm/s. For each thermal desorption experiment, the sample was electrochemically pre-charged with hydrogen according to the process described in Section \ref{Sec:Hcharging}. Upon completion of the cathodic charging, the sample was rinsed with acetone and then distilled water, carefully dried with warm air, and then loaded into the transfer chamber of the TDS. The transfer system was then evacuated to approximately 10$^{-7}$ mbar. After the transfer from the loading chamber was complete, the specimen was placed onto an aluminium nitride (AlN) sheet in direct contact with the controllable, 2.5 kW heating stage. Four ceramic rods were then extended to press the sample against the AlN sheet, thereby maximizing conduction as well as ensuring homogeneous heating of the entire specimen. Once the specimen had been sufficiently fixed against the AlN plate, the desorption experiment was initiated using a pre-programmed temperature versus time sequence. The typical elapsed time between the completion of cathodic charging and the beginning of the thermal desorption experiment was 30 minutes. A schematic illustrating this experimental setup is shown in Fig. \ref{fig:TDSSetup}.

Two types of thermal desorption experiments were performed in the current study to assess the hydrogen diffusivity of cold-rolled pure iron, as described in the following sections.

\subsubsection{Desorbed hydrogen tests}
A series of 10 mm x 10 mm x 1 mm specimens were cathodically pre-charged in 3 wt. \% NaCl solution using a current density of 5 mA/cm$^2$ during a charging time of 3 hours. After charging, the samples were then allowed to rest at ambient temperature ($\approx 22^\circ$C) in laboratory air for times (which included the TDS pump-down time) ranging from 0.5 to 48 hours. Each sample was then heated from 25 to 850$^\circ$C at a rate of 30$^\circ$C/min and the total remaining hydrogen content was calculated \emph{via} integration of the obtained thermal desorption spectra. The hydrogen diffusivity at ambient temperature, $D_{des}$, can then be determined by comparing the experimental hydrogen content versus desorption time relationship to that predicted by published analytical solutions for 1-D diffusion in a plate geometry of half-thickness $L$, such as \citep{Crank1979}:
\begin{equation}\label{eq:Ddes}
    \frac{C-C_0}{C_1-C_0}=1-\frac{4}{\pi}\sum_{n=0}^{\infty}\frac{(-1)^n}{2n+1}\exp\left\lbrace-D_{des}(2n+1)^2\pi^2t/4L^2\right\rbrace \cos\frac{(2n+1)\pi x}{2L}
\end{equation}
%Note that C_0 is the initial concentration and C_1 is the surface concentration.
\noindent where $C$ is the hydrogen concentration at position $x$ after desorption time $t$, $C_0$ is the initial uniformly distributed hydrogen concentration at $t$ = 0 seconds, and $C_1$ is the surface hydrogen concentration. A Matlab script was used to iteratively increment $C_0$ and $D_{des}$ to obtain the optimal fit between the average concentration across the simulated specimen thickness and the measured remaining total hydrogen content for each desorption time. Note that there are several critical assumptions in this analysis. First, given that the desorbed hydrogen content contains the trapped hydrogen, which is unlikely to desorb at ambient temperature, the total hydrogen content was found to asymptotically approach a lower-bound plateau value at long desorption times. This was phenomenologically captured in this simulation by considering $C_1$ to be the trapped hydrogen concentration which, based on the desorption experiment results (see Fig. \ref{fig:DesorptionCurve}), was set to a constant value of 0.75 weight parts per million (wppm). Second, a constant specimen thickness of 1 mm was assumed for all simulations. Lastly, an initially uniform concentration is assumed, in accordance with the boundary conditions used for the derivation of Eq. (\ref{eq:Ddes}) \citep{Crank1979}.

\subsubsection{Isothermal TDS tests}

Isothermal desorption experiments were performed at 22, 40, 60, and 80$^\circ$C on a series of 10 mm x 10 mm x 1 mm specimens that were each cathodically pre-charged in 3 wt. \% NaCl solution using a current density of 5 mA/cm$^2$ during charging time of 3 hours. Duplicate experiments were also performed at 22$^\circ$C and 80$^\circ$C. These isothermal experiments are nominally similar to typical programmed-temperature desorption \citep{Verbeken2012}, but hydrogen egress is monitored under a fixed temperature, as opposed to ramping the temperature versus time profile. For all tested temperatures greater than 22$^\circ$C, the time required to stabilize the sample at the test temperature varied from 140 to 180 seconds.\\

Given the isothermal nature of these experiments, it is straightforward to determine the hydrogen diffusivity from the desorbed hydrogen content versus time profile using numerical analysis approaches (e.g., finite differences or finite element). In this work, we chose to use finite element modelling to determine the hydrogen diffusivity from the isothermal TDS tests. The magnitude of the hydrogen diffusion coefficient $D$ can be determined by fitting the experimental desorption curve with the output of a 1D finite element (FE) simulation of hydrogen transport. Specifically, diffusion is governed by Fick's second law, which in a one-dimensional form reads:
\begin{equation}
    \frac{\partial C}{\partial t} = D \frac{\partial^2 C}{\partial x^2}
\end{equation}

\noindent where $C$ is the diffusible hydrogen concentration. One must also define an initial condition ($C=C_0$ at $t=0$) and suitable boundary conditions ($C=0$ at $x=\pm L/2$). The simulations were then completed as follows. First, the hydrogen egress during experiments associated with the resting time required for the chamber pump-down (30 min.) and subsequent heating time to reach the targeted temperature, $T_{iso}$, was simulated by assuming the sample was held at room temperature ($T=22^\circ$C) for the entire rest and heating duration. Then, upon completing this initial hold time, the temperature in the simulation was instantly increased to $T_{iso}$ and then held constant for the duration of the simulation. The diffusivity used in the initial resting step of the simulations was determined by first fitting the experimentally-measured desorbed hydrogen content versus time data obtained at 22$^\circ$C \textit{via} iteratively changing the $C_0$ and $D$ values. Once this value was known, the best $D$ at $T=22^\circ$C was then applied for the resting period for the higher temperature experiments, with $C_0$ and $D$ then adjusted to fit the observed desorption data for each respectve $T_{iso}$. A sensitivity analysis revealed the existence of a unique pair of $C_0$ and $D$ values that yielded the best fit of the experimental curve for each temperature. Critically, while the simulation results were found to be minorly affected by changes in $C_0$, they were strongly sensitive to subtle changes in $D$, thereby indicating the uniqueness of the best fit $D_{iso}$ for each temperature.

\section{Results}
\label{Sec:Results}

\subsection{Influence of the charging current}
\begin{figure}[H]
     \centering
         \centering
         \includegraphics[width=0.8\textwidth]{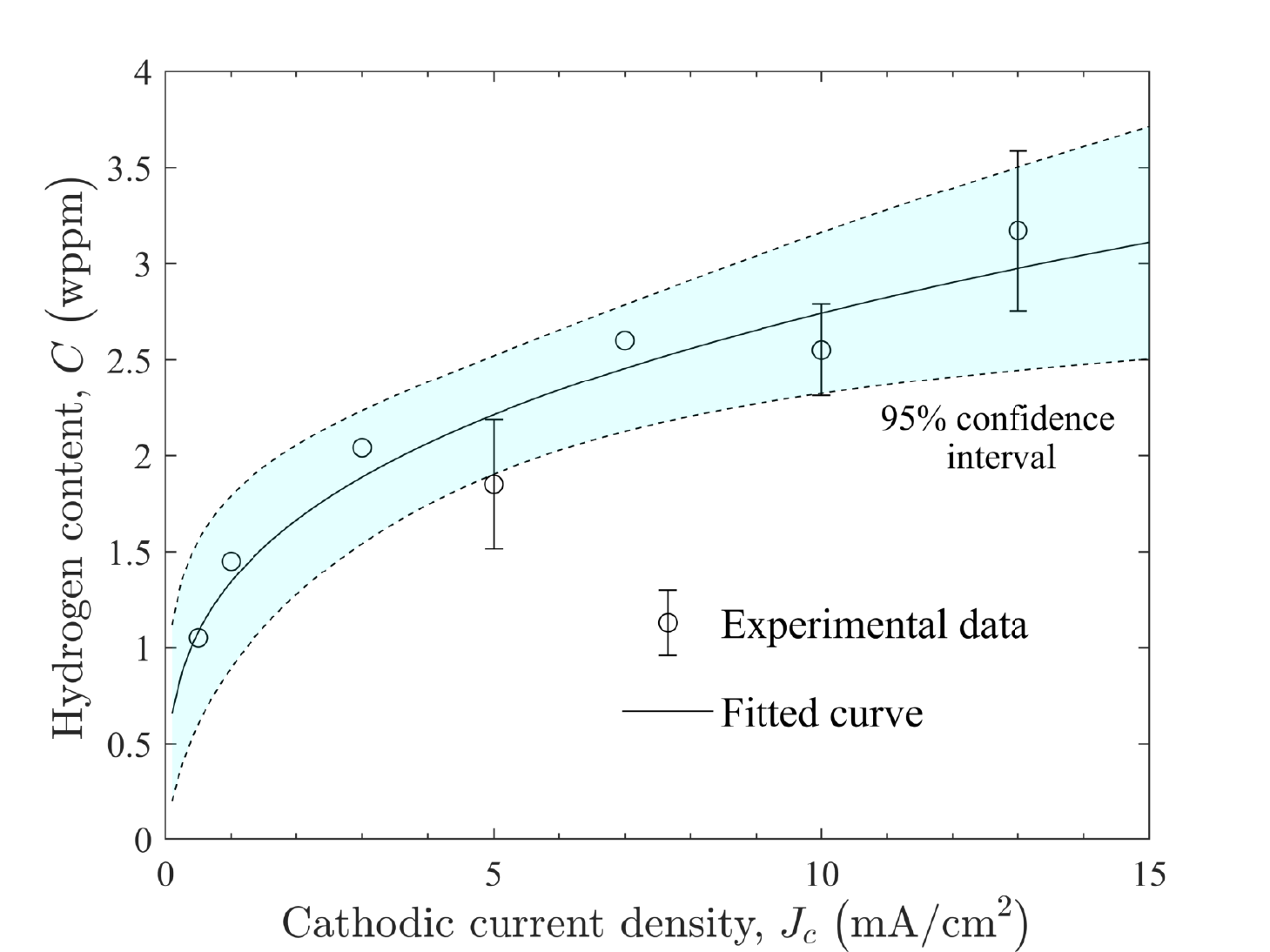}
        \caption{Evolution of the hydrogen concentration with the current density and prediction model.}
        \label{fig:HVScurrentFit}
\end{figure}

First, it is necessary to determine the cathodic current density, $J_c$, to be employed for subsequent hydrogen precharging and permeation experiments, as well as to establish a general relationship between applied current density and total hydrogen content for cold-rolled Fe. Towards this end, specimens were charged at seven different current densities ranging from 0.5 to 13 mA/cm$^2$. The total hydrogen content was then calculated \textit{via} integration of the TDS spectra collected for each specimen over the temperature range of 25 to 850$^\circ$C. A total of 31 specimens were evaluated: 1 specimen at 0.5 mA/cm$^2$, 1 specimen at 1 mA/cm$^2$, 1 specimen at 3 mA/cm$^2$, 22 specimens at 5 mA/cm$^2$, 1 specimen at 7 mA/cm$^2$, 3 specimens at 10 mA/cm$^2$, and 2 specimens at 13 mA/cm$^2$. The measured total hydrogen content $C$ as a function of $J_c$ is shown in Fig. \ref{fig:HVScurrentFit}. For the current densities where multiple experiments were performed, the average measured concentration is plotted, with the error bars representing the calculated standard deviation. These data were then fit to a power law function, yielding the following relationship: $C=1.342J_c^{0.3102}$, as indicated by the solid line in Fig. \ref{fig:HVScurrentFit}. The 95\% prediction bands, which represent the upper and lower bounds between which there is 95\% confidence the fitted function will reside, are represented by dashed lines in Fig. \ref{fig:HVScurrentFit}.

Examination of Fig. \ref{fig:HVScurrentFit} reveals four observations. First, the total hydrogen content exhibits an initially rapid increase with increasing $J_c$ up to 3 mA/cm$^2$, followed by a more gradual increase over the remaining range of evaluated current densities. This asymptotic-like behavior at the relatively low hydrogen contents of between 3 to 3.5 wppm is consistent with the generally low hydrogen solubility of $\alpha$-iron \citep{Choo1981}. Second, regarding the solubility of hydrogen in pure Fe, the literature establishes that it is on the order of 1 wppm \citep{Choo1981}, which is lower than the values reported here. However, it should be noted that the evaluated pure Fe was cold rolled down to its final thickness of 1 mm. As such, an increase in solubility is expected due to the multiplication of defects, which will act as trap sites and increase the total hydrogen concentration \citep{Kiuchi1983,Oriani1970}. Third, as demonstrated by both the upper and lower bounds of the prediction interval and the error bars calculated from duplicate experiments, the dataset is well described by the fitted power law equation. Lastly, the observed relative scatter of the data, especially for the 5 mA/cm$^2$ condition where 22 specimens were evaluated, is in line with expectations from other electrochemical charging studies of hydrogen content \citep{Ai2013,RinconTroconis2017,Harris2021a} and suggests reasonable repeatability amongst the duplicate specimens.

\subsection{Permeation tests}

\begin{figure}[H]
     \centering
         \centering
         \includegraphics[width=0.8\textwidth]{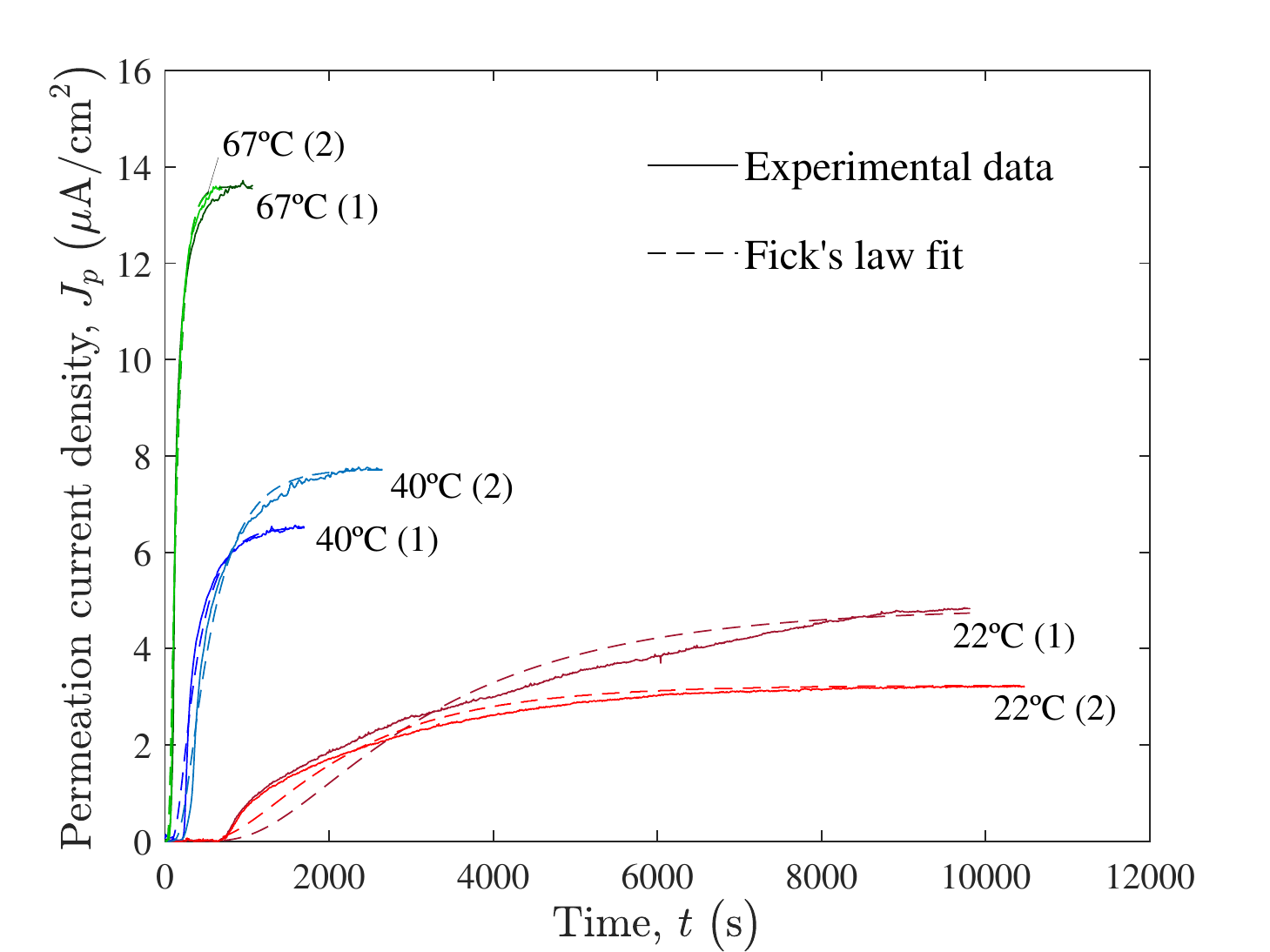}
        \caption{Permeation transients obtained in pure Fe at different temperatures, including a fit using Fick's law.}
        \label{fig:PermeCurvesPlot}
\end{figure}

The permeation curves obtained for duplicate experiments on cold-rolled pure Fe at 22, 40, and 67$^\circ$C, along with their corresponding fits to Eq. (\ref{eq:Dlpc}), are shown in Fig. \ref{fig:PermeCurvesPlot}. For each experiment, the steady state permeation current density ($J_\infty$), breakthrough time ($t_{bt}$), and lag time ($t_{lag}$) are documented in Table \ref{Permeation}. The calculated hydrogen diffusivities using the three approaches described in Section \ref{Subsec:Electropermeation} are also noted. First, considering the results at 22$^\circ$C, the diffusivity is found to  vary between 4 and $7.4\times10^{-11}$ m$^2$/s across the three employed methods, which is broadly consistent with reported hydrogen diffusivities for cold-rolled pure Fe \citep{VandenEeckhout2017}. Second, both the slope of the permeation transient and the magnitude of $J_\infty$ systematically increase with temperature, as expected from literature results \citep{Addach2005,Barrer1940a}. This implies an increase of the hydrogen diffusivity with temperature, as can be seen in Table \ref{Permeation}, across all the methodologies employed. Third, it is important to note that the assumed diffusivity employed to estimate the time required for hydrogen precharging of the desorption experiments ($D \approx 8\times10^{-11}$ m$^2$/s; \citealp{VandenEeckhout2017}) was a reasonable estimate. Critically, considering the calculated diffusivities for 22$^\circ$C in Table \ref{Permeation}, it is expected that the centerline of the 1-mm thick specimens used in the current study would indeed be nominally saturated after 3 hours; \textit{i.e.}, reached between 98.20$\%$ and 99.95$\%$ of the surface hydrogen concentration, as per Eq. (\ref{eq:Ddes}).  

\begin{table}[H]
 \centering
  \caption{Diffusivity values calculated from the permeation transients.}
  \label{Permeation}
\begin{adjustwidth}{-0.5cm}{-0.5cm}
\begin{tabular}[t]{cccccccc}
\toprule
\text{Test} & \text{$J_\infty$ ($\mu$A/cm$^2$)} & \text{$t_{bt}$(s)} & \text{$D_{bt}$ (m$^2$/s)} & \text{$t_{lag}$ (s)} & \text{$D_{lag}$ (m$^2$/s)} & \text{$D_{lpc}$ (m$^2$/s)}\\
\midrule
22$^\circ$C (1)  & 4.85 & 870 & 7.36$\times10^{-11}$ & 4110 & 3.97$\times10^{-11}$ & 4.53$\times10^{-11}$  \\
22$^\circ$C (2) & 3.24 & 834 & 6.78$\times10^{-11}$ & 2596 & 5.55$\times10^{-11}$ & 5.89$\times10^{-11}$  \\
40$^\circ$C (1) & 6.52 & 245 & 2.21$\times10^{-10}$ & 377 & 3.66$\times10^{-10}$ & 3.31$\times10^{-10}$  \\
40$^\circ$C (2) & 7.73 & 315 & 1.83$\times10^{-10}$ & 570 & 2.57$\times10^{-10}$ & 2.41$\times10^{-10}$  \\
67$^\circ$C (1) & 13.60 & 87 & 6.22$\times10^{-10}$ & 145 & 9.52$\times10^{-10}$ & 8.52$\times10^{-10}$\\
67$^\circ$C (2) & 13.56 & 80 & 7.19$\times10^{-10}$ & 151 & 9.71$\times10^{-10}$ & 9.24$\times10^{-10}$ \\
\bottomrule
\end{tabular}
\end{adjustwidth}
\end{table}

Examination of the Laplace Fick's law fit for each experiment in Fig. \ref{fig:PermeCurvesPlot} demonstrates that steady state conditions were attained in each experiment, as shown by the good agreement between the experimental and fitted curves. The exception is the 22$^\circ$C (1) case, where  stabilisation of the permeation curve was not observed. Speculatively, the difficulty in reaching steady state for this experiment may be due either to surface effects taking place in the cathodic or anodic sides of the specimen \citep{Kiuchi1983,Addach2009} or to incomplete filling of the hydrogen traps \citep{Kumnick1974,Oriani1970}. As such, it is likely that the reported diffusivities determined for this test are biased to slightly higher values. This situation exemplifies a primary limitation of electropermeation tests when assessing hydrogen diffusivity in metals as numerous variables can hinder the attainment of steady state, thereby obfuscating subsequent analysis efforts.

Comparing the calculated diffusivities for each method in Table \ref{Permeation}, it is interesting to note that the lag time and the Laplace methods resulted in similar hydrogen diffusivities across all tested specimens, with the largest differences being less than 12\%. Conversely, the diffusivities determined with the breakthrough method at 22$^\circ$C were noticeably higher than those obtained with the lag time and Laplace methods, but then found to be lower in the case of the tests performed at higher temperatures. Speculatively, this behaviour could be explained in terms of the progressive loss of importance of trapping phenomena over lattice diffusion as the temperature is increased \citep{Oriani1970}. It is worth noting that the differences in diffusivity between the breakthrough and the lag time/Laplace methods increased to nearly 40\% in some cases, which is greater than the scatter reported between duplicate experiments.

\subsection{Desorption curve}

\begin{figure}[H]
     \centering
         \centering
         \includegraphics[width=0.8\textwidth]{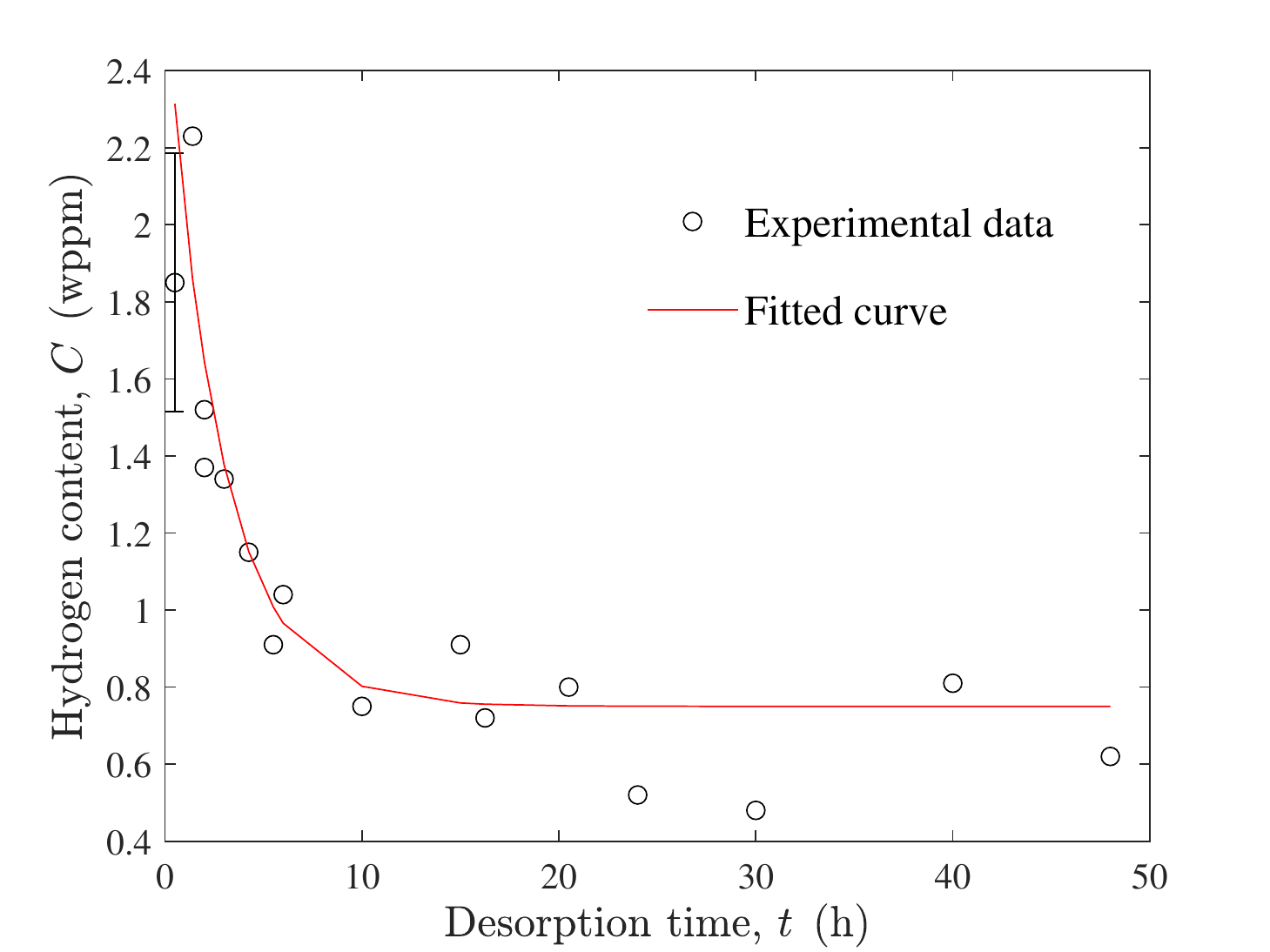}
        \caption{Desorption curve. Hydrogen concentration as a function of elapsed time in air at room temperature. The datapoint and error bars at 0.5 hours represent the average and standard deviation of the total content measured from 22 experiments.}
        \label{fig:DesorptionCurve}
\end{figure}

The results of the `rest time'-based approach to measuring hydrogen diffusivity \textit{via} desorption methods is shown in Fig. \ref{fig:DesorptionCurve}. As expected, the total hydrogen concentration within the specimen exhibits an initially strong reduction with increasing rest time at ambient temperature, eventually leveling out at a nominal concentration of 0.75 wppm after approximately 10-15 hours. This initial steep reduction is consistent with the egress of the diffusible hydrogen concentration, while the observed plateau is indicative of the trapped hydrogen content. These data were then iteratively fit to Eq. (\ref{eq:Ddes}), with $C_0$ and $D_{des}$ used as fitting parameters. The results of the best obtained fit are indicated by the red line in Fig. \ref{fig:DesorptionCurve}, which corresponded to $C_0$ = 3.0 wppm and $D_{des} = 1\times10^{-11}$ m$^2$/s. Comparing this value to those obtained from the various permeation methods at 22$^\circ$C (Table \ref{Permeation}), it is clear that the `rest time' desorption approach yields a tangibly lower diffusivity. Possible reasons for this decreased diffusivity will be enumerated in the Discussion section.

\subsection{Isothermal TDS tests}

\begin{figure}[H]
     \centering
         \centering
         \includegraphics[width=0.85\textwidth]{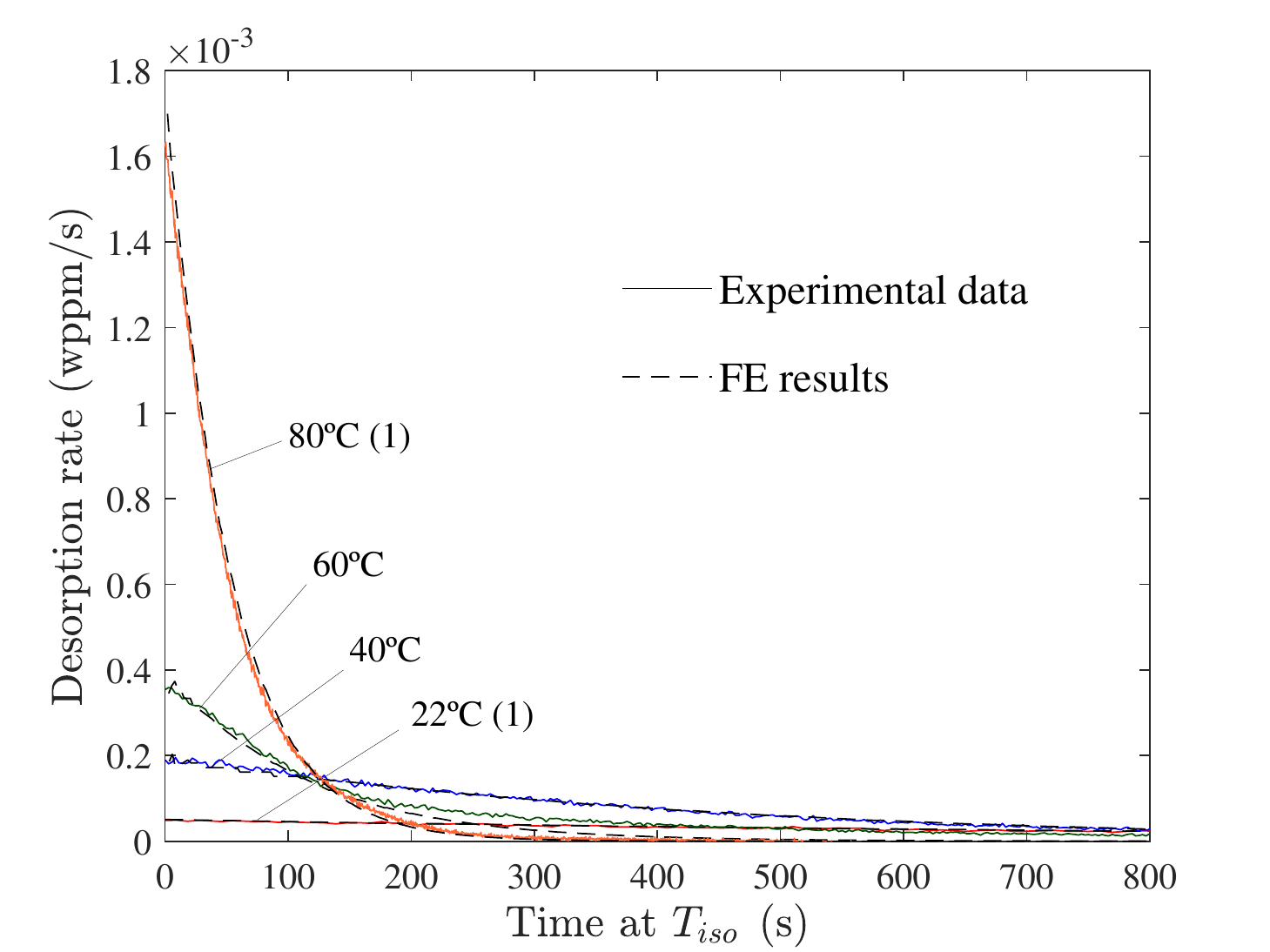}
        \caption{Isothermal TDS tests (solid lines) performed at different temperatures and results from the finite element analysis to estimate $D$ (dashed lines).}
        \label{fig:IsothermalAllFit}
\end{figure}

The measured desorption rate versus time profiles during isothermal desorption experiments at temperatures of 22, 40, 60 and 80$^\circ$C are shown in Fig. \ref{fig:IsothermalAllFit}. The initial slope of the desorption rate versus time relationship is noted to increase with increasing temperature, consistent with expectations for an increasing hydrogen diffusivity and with the behavior observed during the permeation experiments (Fig. \ref{fig:PermeCurvesPlot}). The best fit to Fick's second law, determined from finite element (FE) simulations, is indicated by the corresponding dashed lines for each respective condition. For all cases, the FE calculations closely capture the experimentally observed initial slope, which is a strong function of the hydrogen diffusivity. The initial diffusible hydrogen concentration ($C_{0}$) and diffusion coefficient ($D_{iso}$) that yielded the best fit of the experimental curve with Fick's second law (FE model) are shown in Table \ref{Isothermal}. As expected, and in line with the permeation results, the calculated diffusivity exhibits a systematic increase with increasing temperature. 

Finally, it is worth noting that very similar diffusion coefficients were obtained for the replicate tests performed at 22$^\circ$C and 80$^\circ$C, respectively. This is observed even though noticeable differences are seen in the fitted $C_{0}$ values, though these variations in $C_{0}$ lie within the scattering reported in Fig. \ref{fig:HVScurrentFit}. This excellent repeatability suggests this method - isothermal TDS tests combined with numerical fitting - is a robust approach for the determination of hydrogen diffusivities.

\begin{table}[H]
 \centering
  \caption{Initial diffusible hydrogen concentration, $C_{0}$, and diffusivity values, $D_{iso}$, estimated through FE modelling from isothermal TDS tests.}
  \label{Isothermal}
\begin{tabular}[t]{cccccccc}
\toprule
\text{Test} & \text{$C_{0}$ (wppm)} & \text{$D_{iso}$ (m$^2$/s)}\\
\midrule
22$^\circ$C (1) & 0.27 & $7.8\times10^{-11}$ \\
22$^\circ$C (2) & 0.45 & $7.5\times10^{-11}$ \\
40$^\circ$C & 0.70 & $2.0\times10^{-10}$ \\
60$^\circ$C & 0.42 & $7.0\times10^{-10}$ \\
80$^\circ$C (1) & 0.68 & $1.7\times10^{-9}$ \\
80$^\circ$C (2) & 0.19 & $1.3\times10^{-9}$ \\
\bottomrule
\end{tabular}
\end{table}

\section{Discussion}
\label{Sec:Discussion}

\begin{figure}[H]
     \centering
         \centering
         \includegraphics[width=0.8\textwidth]{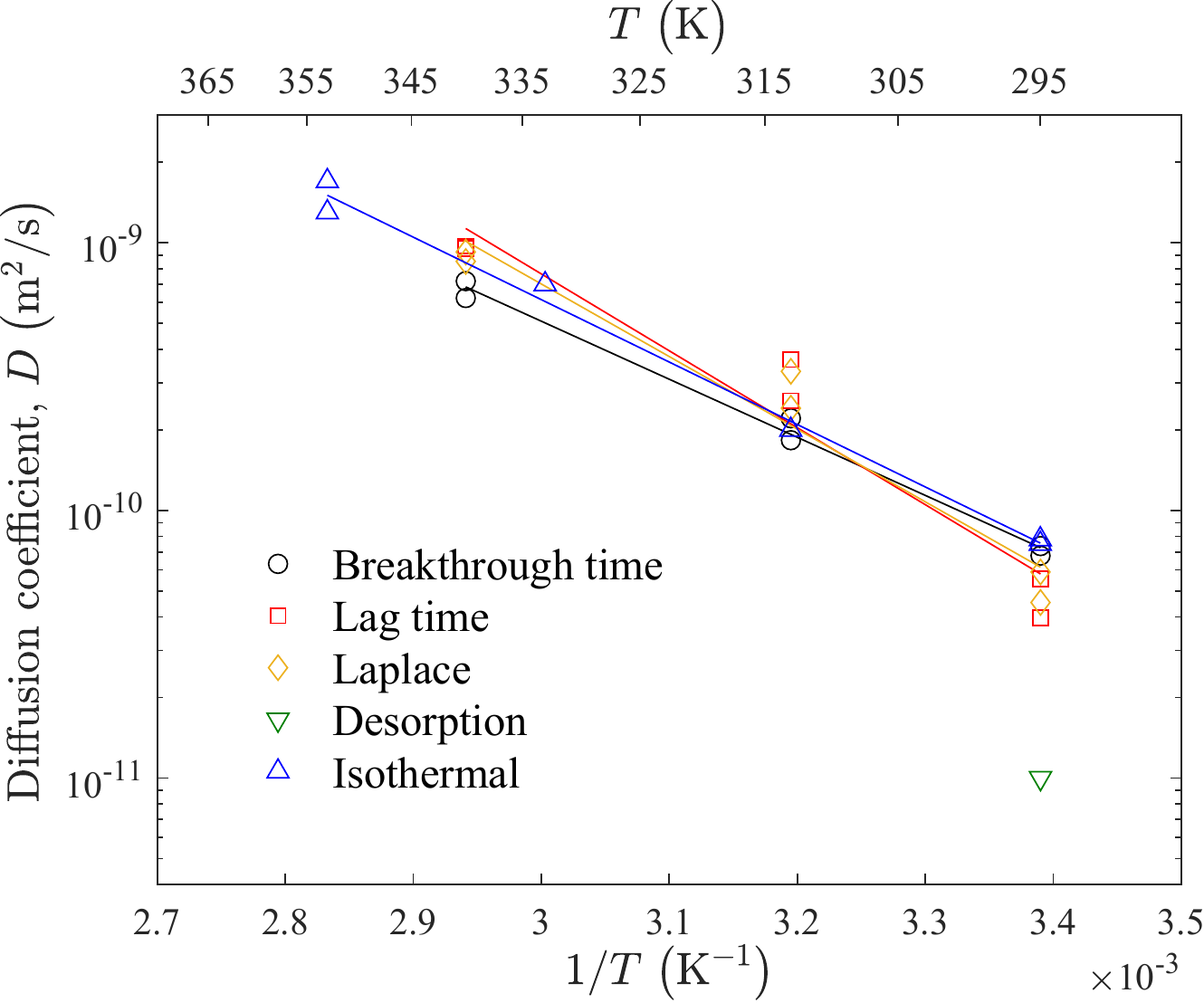}
        \caption{Estimated diffusivities versus the inverse of temperature for the various permeation and desorption methods investigated.}
        \label{fig:ArrheniusLinearIso}
\end{figure}

We proceed to compare and analyse the results obtained by the various permeation and desorption methods. The estimated diffusivities are shown versus the inverse of temperature in Fig. \ref{fig:ArrheniusLinearIso}, using a log-linear plot. Here, the solid lines represent a fit of each respective dataset to the following Arrhenius expression:
\begin{equation}\label{eq:Arrhenius1}
    D=D_0 \exp \left(\frac{-E_a}{RT}\right)
\end{equation}
\noindent where $D_0$ is the pre-exponential factor, $E_a$ the activation energy for hydrogen diffusion, and $R$ is the gas constant. The independent term of the linear regression is equal to log($D_0$) and $E_a$ is obtained by multiplying the slope by $R$. The fitted values of $D_0$ and $E_a$ are shown in Table \ref{ArrheniusAll}.

\begin{table}[H]
 \centering
  \caption{Pre-exponential factor, $D_0$, and activation energy for hydrogen diffusion, $E_a$, as determined from permeation and isothermal desorption.}
   \label{ArrheniusAll}
\begin{tabular}[t]{cccccccc}
\toprule
\text{Method} & \text{$D_0$ (m$^2$/s)} & \text{$E_a$ (kJ/mol)}\\
\midrule
Breakthrough time  & $1.65\times10^{-3}$ & 41.5 \\
Lag time & $3.22\times10^{-1}$ & 55.0 \\
Laplace & $9.70\times10^{-2}$ & 51.9 \\
Isothermal & $6.68\times10^{-3}$ & 44.9 \\
\bottomrule
\end{tabular}
\end{table}

Four observations can be made from these data. First, the desorption-based `rest time' method yielded a significantly lower hydrogen diffusivity at 22$^\circ$C (nearly five-fold) as compared to the other four approaches. Given that this approach is based on fitting the measured remaining hydrogen content as a function of outgassing time, it is speculated that variations in absorbed hydrogen concentration are likely responsible for the low observed diffusivity. As demonstrated by the error bars shown in Fig. \ref{fig:DesorptionCurve} for the 22 duplicate measurements performed at 0.5 hours, test-to-test scatter in initially absorbed hydrogen content is non-negligible, which would introduce increased variability into this method relative to other approaches. Second, the isothermal desorption method yielded generally similar diffusivity values for a given temperature relative to the three permeation based approaches. Such agreement between the diffusivity values measured through both methods was not expected, as considerable differences between absorption-based and desorption-based measurements of diffusivity have been previously reported for elevated temperature ($>400^\circ$C)-based evolution experiments in pure iron \citep{Carmichael1960}. Third, examination of the spread in diffusivities at ambient temperature indicates a maximum of two-fold difference across the four best methods (permeation and isothermal desorption), which represents excellent agreement relative to the reported spread in ambient temperature diffusivity for pure Fe in the literature \citep{Ono1968,Kumnick1974,Kiuchi1983}. Lastly, examination of the calculated $E_a$ for the four approaches that were conducted at multiple temperatures demonstrates that the isothermal desorption-based $E_a$ sits in the middle of the three permeation-based approaches. Speculatively, it is likely that the agreement between these different methods would only increase if the isothermal desorption results were compared against the second permeation transient, which represents a scenario where the deep hydrogen traps are already filled.\\

As expected, the diffusion coefficients determined from the current experiments shown in Table \ref{ArrheniusAll} are significantly different from those measured in the literature for well-annealed pure Fe. For example, a thorough analysis of the existing diffusivity literature in well-annealed iron was performed by Kiuchi and McLellan, who suggested that hydrogen diffusion is best described by $D_0=7.23\times10^{-8}$ m$^2$/s and an activation energy $E_a$=5.69 kJ/mol up to a temperature of 80$^\circ$C \citep{Kiuchi1983}. Considering the activation energy parameter, it is expected that the significantly increased dislocation density induced during the cold-rolling process will result in the widespread distribution of modestly strong hydrogen trap sites, thereby increasing the macroscale barrier for hydrogen diffusion \citep{Oriani1970}. For example, \citet{Choo1983} observed a progressive increase in both $D_0$ and $E_a$ with the degree of cold rolling for pure iron. As such, the measured diffusivity is no longer controlled simply by lattice diffusion, but rather becomes dependent on the trapping behavior, resulting in diffusion being best described by an \textit{effective diffusivity} \citep{Oriani1970}.\\

\begin{figure}[H]
     \centering
         \centering
         \includegraphics[width=0.8\textwidth]{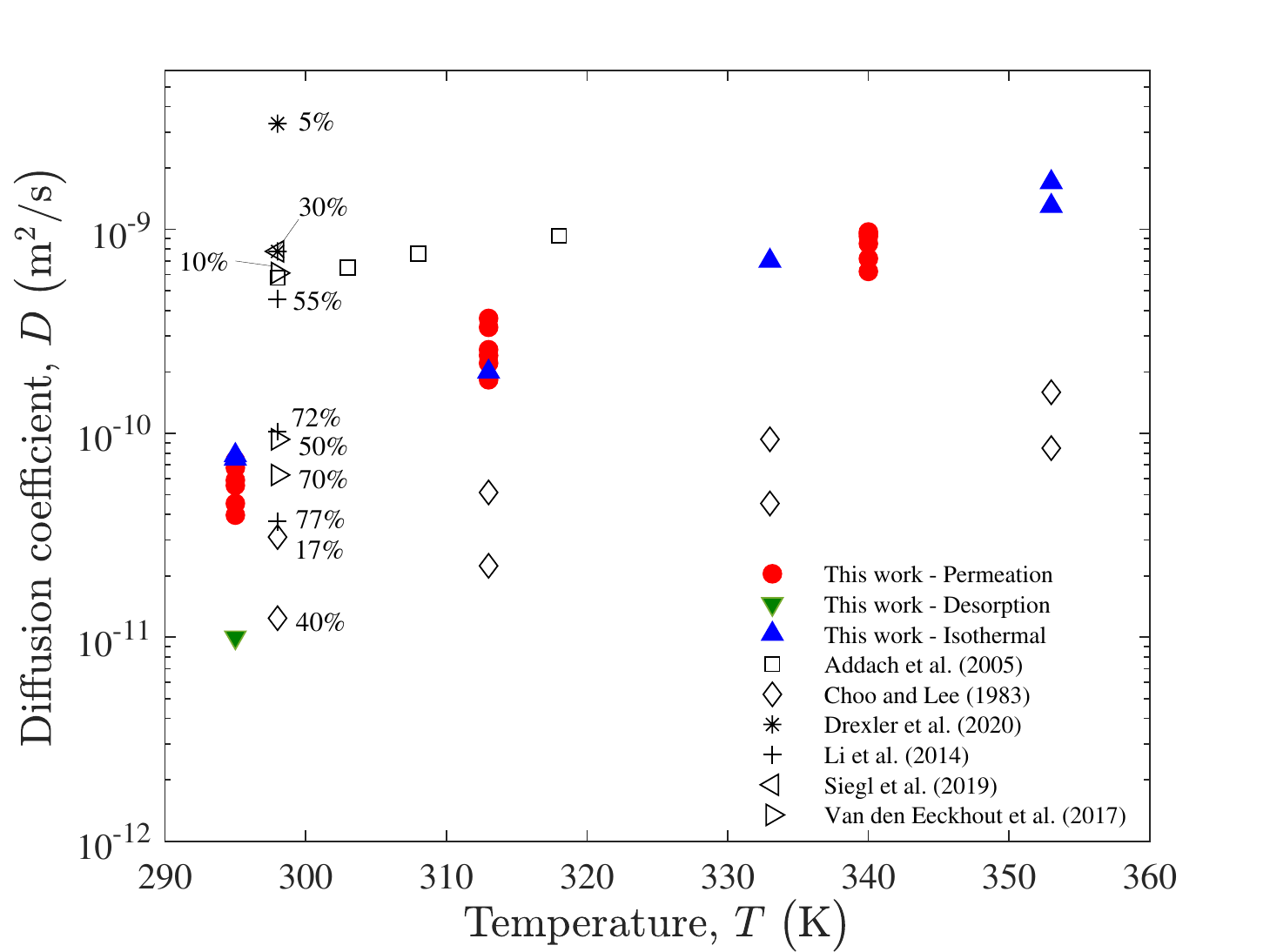}
        \caption{Measured and reported values of the diffusion coefficient of hydrogen in cold-rolled iron as a function of temperature. The degree of cold rolling is indicated; the data from the present work corresponds to a supplier-reported average degree of cold work of 50\%.}
        \label{fig:DvsTDiscusionLiterature}
\end{figure}

Given this expected departure from the behavior of annealed pure Fe, the current study's results should be compared to the diffusivity versus temperature data reported from studies on cold-rolled Fe. This comparison is shown in Fig. \ref{fig:DvsTDiscusionLiterature} for six relevant studies \citep{Addach2005,Choo1983,Drexler2020,Siegl2019,VandenEeckhout2017,Li2014}. The degree of cold rolling involved in each study has been included next to the corresponding data point in Fig. \ref{fig:DvsTDiscusionLiterature}. As was discussed previously based on analyses conducted in the literature \citep{Kiuchi1983}, literature data exhibits substantial scatter in measured diffusivity at ambient temperature (300 K). Fig. \ref{fig:DvsTDiscusionLiterature} demonstrates that this large variability persists in cold-worked iron, as iron with similar degrees of cold work ($\sim$50\%) exhibits one order of magnitude differences in measured diffusivity. However, it is notable that the current results fall approximately in the middle of the observed scatter, suggesting general agreement with these prior data. Less literature data on cold-rolled iron is available at elevated temperatures, but the trend of the current results residing in the nominal middle of the scatter band appears to generally hold as the temperature is increased. 

Collectively, these comparisons with prior literature data and the reasonable agreement with the three different permeation-based approaches (Fig. \ref{fig:ArrheniusLinearIso} and Table \ref{ArrheniusAll}) strongly support the viability of the employed isothermal desorption approach for assessing hydrogen diffusivity in cold-rolled Fe and other `fast' diffusion alloy systems. Moreover, the efficacy of this method was demonstrated using a relatively thin material form (1-mm thick sheet), indicating that the approach does not inherently require the large sample sizes employed in prior isothermal desorption-based studies \citep{Yamabe2015}. Given the large scatter band inherent to permeation-based measurements of the hydrogen diffusivity at low temperatures in pure Fe \citep{Kiuchi1983,Kumnick1974}, the use of alternative approaches should be considered and further investigated. In particular, it would be worthwhile to establish the degree of scatter in isothermal desorption-based diffusivity measurements of cold-rolled Fe. Moreover, as demonstrated by Fig. \ref{fig:DvsTDiscusionLiterature}, the data density of hydrogen diffusivity measurements for cold-rolled Fe significantly decreases as the temperature is increased. The current study provides an initial set of reliable data for such elevated temperatures, but additional studies are needed to further populate the elevated temperature regime.

\section{Conclusions}
\label{Sec:ConcludingRemarks}

The hydrogen diffusivity for cold-rolled pure Fe was measured using both electrochemical permeation and thermal desorption-based techniques, offering an opportunity to compare these approaches. Based on these experiments, the following conclusions were made:
\begin{itemize}
    \item Analysis of permeation data using the breakthrough time method, the lag time method, and a Laplace solution to the entire permeation transient yielded nominally similar hydrogen diffusivities across methodologies for each temperature (22$^\circ$C, 40$^\circ$C and 67$^\circ$C). The hydrogen diffusivity values obtained at 22$^\circ$C were found to be generally consistent with those reported in the literature for cold-rolled pure Fe.
    
    \item Comparison of two desorption-based approaches to calculating the hydrogen diffusion resulted in significantly different values, with differences between isothermal desorption and the rest time method being of almost an order of magnitude. This difference is speculatively attributed to the strong sensitivity of the rest time method to the initial hydrogen concentration in each specimen, which was shown in Fig. \ref{fig:HVScurrentFit} to exhibit substantial variation from test-to-test.
    
    \item Analysis of the Arrhenius relationships constructed from the hydrogen diffusivity versus temperature relationships demonstrates that the isothermal desorption approach yields nominally similar values of diffusion pre-exponential factor ($D_0$) and activation energy ($E_a$) to the permeation breakthrough time method. Slight differences are observed when comparing the isothermal results to lag time and Laplace permeation approaches. However, a collective analysis suggests that isothermal desorption is generally consistent with the permeation results.
    
    \item Hydrogen diffusivity values reported in the literature as a function of temperature for cold-rolled Fe are in excellent agreement with those determined in the current study using permeation and isothermal desorption.
    
    \item The results of this study demonstrate the efficacy of the isothermal desorption technique for evaluating hydrogen diffusivity in thin-section specimens of `fast' diffusing materials. Additional studies are needed to evaluate the experimental scatter of this approach relative to that observed for permeation-based experiments, but initial results suggest that isothermal desorption exhibits reduced levels of scatter.
\end{itemize}

\section{Acknowledgements}
\label{Sec:Acknowledgeoffunding}

Emilio Mart\'{\i}nez-Pa\~neda acknowledges discussions with A. D\'{\i}az (University of Burgos). The authors acknowledge financial support from the EPSRC (grants EP/V04902X/1 and EP/V009680/1).

%\appendix

%\section{Additional details of numerical implementation}
%\label{App:FEM}

% Appendix A

%% If you have bibdatabase file and want bibtex to generate the
%% bibitems, please use
%%
%%  \bibliographystyle{elsarticle-harv} 
%%  \bibliography{<your bibdatabase>}

%% else use the following coding to input the bibitems directly in the
%% TeX file.

%\bibliographystyle{elsarticle-num}
\bibliographystyle{elsarticle-harv}
\bibliography{library}

%% \bibitem[Author(year)]{label}
%% Text of bib
\end{document}